\documentclass[aps,pra,10pt,twocolumn,superscriptaddress]{revtex4-1}
\usepackage{amsmath}
\usepackage{amssymb}
\usepackage{bbm}
\usepackage{hyperref}
\usepackage{graphicx}
\usepackage{framed}

\newlength{\eqboxstorage}
\newcommand{\eqbox}[1]{
\setlength{\eqboxstorage}{\fboxsep}
\setlength{\fboxsep}{6pt}
\boxed{#1}
\setlength{\fboxsep}{\eqboxstorage}
}

\begin{document}

\title{Experiments testing macroscopic quantum superpositions must be slow}

\author{Andrea Mari}
\affiliation{NEST, Scuola Normale Superiore and Istituto Nanoscienze-CNR, I-56126 Pisa, Italy}
\author{Giacomo De Palma}
\affiliation{NEST, Scuola Normale Superiore and Istituto Nanoscienze-CNR, I-56126 Pisa, Italy}
\affiliation{INFN, Pisa, Italy}
\author{Vittorio Giovannetti}
\affiliation{NEST, Scuola Normale Superiore and Istituto Nanoscienze-CNR, I-56126 Pisa, Italy}

\begin{abstract}
We consider a thought experiment where the preparation of a macroscopically massive or charged particle in a quantum superposition and the associated dynamics of a distant test particle apparently allow for superluminal communication.
We give a solution to the paradox which is based on the  following fundamental principle: any local experiment, discriminating a coherent superposition from an incoherent statistical mixture, necessarily requires a minimum time proportional to the mass (or charge) of the system.
For a charged particle, we consider two examples of such experiments, and show that they are both consistent with the previous limitation.
In the first, the measurement requires to accelerate the charge, that can entangle with the emitted photons.
In the second, the limitation can be ascribed to the quantum vacuum fluctuations of the electromagnetic field.
On the other hand, when applied to massive particles our result provides an indirect evidence for the existence of gravitational vacuum fluctuations and for the possibility of entangling a particle with quantum gravitational radiation.
\end{abstract}
\maketitle

\section{Introduction}

The existence of coherent superpositions is a fundamental postulate of quantum mechanics but, apparently, implies very counterintuitive consequences when extended to macroscopic systems. This problem, already pointed out since the beginning of quantum theory through the famous Schr\"odinger cat paradox \cite{wheeler}, has been the subject of a large scientific debate which is still open and very active.

Nowadays there is no doubt about the existence of quantum superpositions.
Indeed this effect has been demonstrated in a number of experiments involving microscopic systems (photons \cite{taylor,dopfer}, electrons \cite{donati, tonomura}, neutrons \cite{zeilinger}, atoms \cite{anderson, monroe}, molecules \cite{arndt, eibenberger}, {\it etc.}).
However, at least in principle, the standard theory of quantum mechanics is valid at any scale and does not put any limit on the size of the system:  if you can delocalize a molecule then nothing should forbid you to delocalize a cat, apart from technical difficulties. Such difficulties are usually associated with the impossibility of isolating the system from its environment, because it is well known that any weak interaction changing the state of the environment is sufficient to destroy the initial coherence of the system.

In this work we are interested in the ideal situation in which we have a macroscopic mass or a macroscopic charge perfectly isolated from the environment and prepared in a quantum superposition of two spatially separated states. Without using  any speculative theory of quantum gravity or sophisticated tools of quantum field theory, we propose a simple thought experiment based on  particles interacting via semiclassical forces. Surprisingly a simple consistency argument with relativistic causality is enough to obtain a fundamental result which, being related to gravitational and electric fields, indirectly tells us something about quantum gravity and quantum field theory.

\begin{figure}[tr]
\includegraphics[width= \columnwidth]{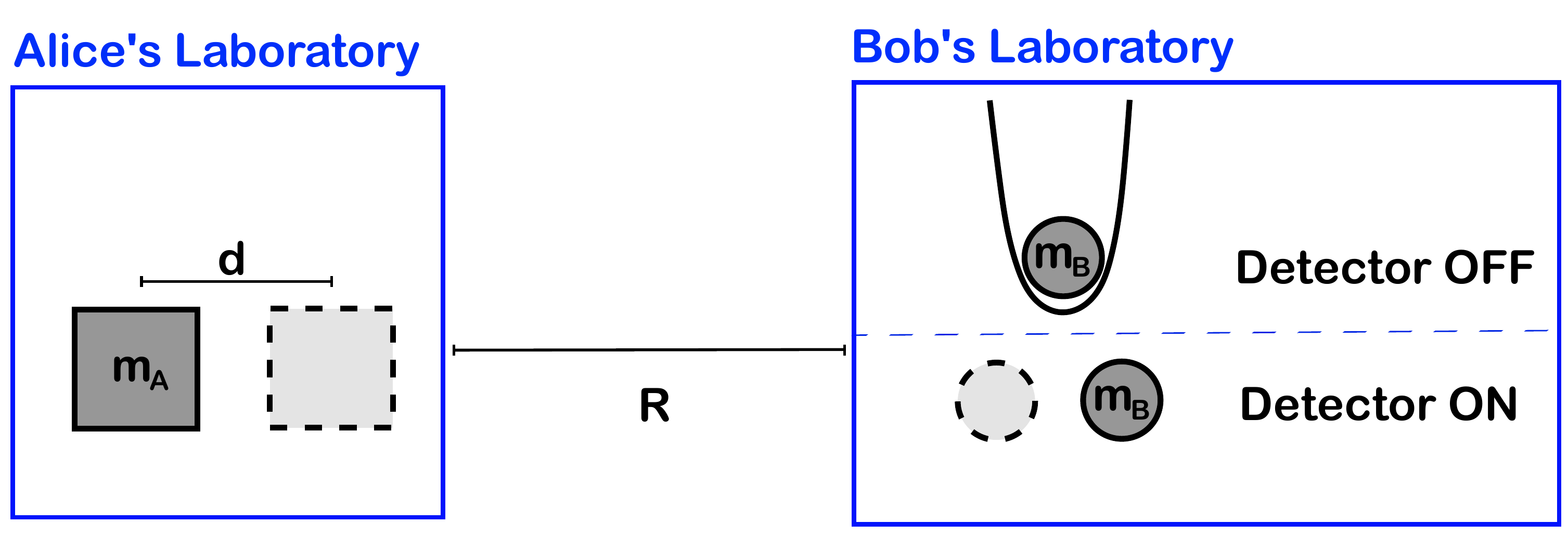}
\caption{Picture of the thought experiment. Alice prepares a macroscopic mass in a quantum spatial superposition. Bob has at disposal a test mass prepared in the ground state of a narrow harmonic trap. Bob can send one bit of information to Alice by choosing between two alternatives: doing nothing (detector {\it off}) or removing the trap (detector {\it on}). Once a time $T_{\mathrm{B}}$ necessary to generate entanglement (if the detector is {\it on}) has passed,
Alice performs a measurement in a time $T_{\mathrm{A}}$ in order to discriminate the coherent superposition from a classical incoherent mixture. In this way, by knowing whether the detector is {\it on} or {\it off}, she gets the information sent by Bob in a time $T_{\mathrm{A}}+T_{\mathrm{B}}$. A completely  equivalent protocol can be obtained by replacing massive particles with charged particles.   } \label{gedanken}
\end{figure}

The result is the following: assuming that a macroscopic mass $m$ is prepared in a superposition of two states separated by a distance $d$, then any experiment discriminating the coherent superposition from a classical incoherent mixture requires a minimum time $T \propto m \, d$, proportional to the mass and the separation distance. Analogously for a quantum superposition of a macroscopic charge $q$, such minimum time is proportional to the associated electric dipole  $T \propto q\, d$. In a nutshell, experiments testing macroscopic superpositions are possible in principle, but they need to be slow.
For common experiments  involving systems below the Planck mass and the Planck charge this limitation is irrelevant, however such time can become very important  at macroscopic scales. As an extreme example, if the center of mass of the Earth were in a quantum superposition with a separation distance of one micrometer, according to our result  one would need a time equal to the age of the universe in order to distinguish this state from
a classical statistical mixture. Clearly this limitation suggests that at sufficiently macroscopic scales quantum mechanics can be safely replaced by classical statistical mechanics without noticing the difference.

 The fact that large gravitational or electromagnetic fields can be a limitation for the observation of quantum superpositions is not a new idea. In the past decades, several models of spontaneous localization \cite{penrose, GRW, diosi,karolyhazy, review} have been proposed which, going beyond the standard theory of quantum mechanics,  postulate the existence a gravity induced collapse at macroscopic scales. Remaining within the domain of standard quantum mechanics, the loss of coherence in interference experiments due to the emission of electromagnetic radiation has been already studied in the literature \cite{petruccione,baym}. Similarly, the interaction of a massive particle with gravitational waves \cite{blencowe,suzuki,jaekel} and the dephasing effect of time dilation on internal degrees of freedom \cite{brukner} have been considered as possible origins of quantum decoherence.

  For what concerns our thought experiment, a similar setup can be found in the literature where the interference pattern of an electron passing through a double slit is destroyed by a distant measurement of its electric field. This thought experiment can be traced back to Bohr as quoted in \cite{baym},  was discussed by Hardy interviewed in \cite{schlosshauer} and appears as an exercise in the book by Aharanov and Rohrlich \cite{aharanov}. Moreover, recently different experiments involving interacting test particles have been proposed in order to discriminate the quantum nature of the gravitational field from a potentially classical description \cite{kafri,bahrami}.

The  original contribution of our work is that, imposing the consistency with relativistic causality, our thought experiment  allows the derivation of a fundamental minimum time which is valid for {\it any} possible experiment involving macroscopic superpositions. In this sense our bounds represent universal limitations having a role analogous to the Heisenberg uncertainty principle in quantum mechanics. For this reason, while our results could be observable in advanced and specific experimental setups \cite{romero-isart, marshall, pikovski, bawaj,schnabel,scala,wan}, their main contribution  is probably a better understanding of the theory of quantum mechanics at macroscopic scales.
For charged particles we propose two different measurements for testing the coherence.
The first requires to accelerate the charge, and our bound on the discrimination time is due to the entanglement with the emitted photons.
In the second, the bound can be instead ascribed to the presence of the vacuum fluctuations of the electromagnetic field.
On the other hand we also find an equivalent bound associated to quantum superposition of large masses.
What is the origin of this limitation?
The analogy suggests that the validity of our bound could be interpreted as an indirect evidence for the existence of  quantum fluctuations of the gravitational field, and of quantum gravitational radiation.

This work is structured in the following way: in section~\ref{experiment} we propose our thought experiment which suggests a minimum discrimination time for any macroscopic quantum superposition. In section~\ref{time} we derive a quantitative bound. Finally in sections~\ref{ent} and~\ref{QED} we check the consistency of our results with an explicit analysis of a charged particle interacting with the electromagnetic field.
Here we propose two different measurements, and show that in both cases they are able to check the coherence only if their duration satisfies our fundamental limit.

\section{Thought  experiment	}
\label{experiment}

Consider the thought experiment represented in Fig.\ \ref{gedanken}, and described by the following protocol. The protocol can be equivalently applied to quantum superpositions of large masses or large charges.

\begin{framed}
\noindent {\bf {Protocol of the thought experiment}}
\begin{enumerate}
\item Alice has at disposal, in her laboratory, a massive/charged particle in a macroscopic superposition of  a ``left'' and  a ``right'' state:
\begin{equation}\label{LR}
 | \psi \rangle =\frac{ | {\rm L} \rangle + | {\rm R}\rangle}{\sqrt{2}}.
\end{equation}
The wave functions of the two states are $\langle x | {\rm L}  \rangle = \phi(x)$ and $\langle x | {\rm R}  \rangle = \phi(x-d)$, where  $d>0$ is the relative separation of the superposition.
\item Bob is in a laboratory at a distance $R$ from Alice and containing a charged/massive test particle prepared in the ground state of a very narrow harmonic trap.  Bob freely chooses between two options: doing nothing ( detector = {\it off}), or removing the trap (detector = {\it on}). In the first case the state of test particle remains unchanged while, in the second case,  the dynamics is sensitive to the local Newton/Coulomb field generated by Alice's particle and the global state will eventually become entangled. If the detector is {\it off}, the initial quantum superposition is preserved, while if the detector is {\it on} the generation of entanglement eventually destroys the coherence of the reduced state of Alice.
\item Alice performs an arbitrary measurement in her laboratory with the task of discriminating the coherent superposition from a statistical incoherent mixture of the two states $|{\rm L}  \rangle$ and $|{\rm R}  \rangle$. For example she could make an interference experiment, a measurement of the velocity, a projection on the symmetric/antisymmetric basis $|{\rm L} \rangle \pm | {\rm R} \rangle $, {\it etc.}. The specific details of the experiment are irrelevant. Depending on the result of the measurement, Alice deduces the choice of  Bob ({\it i.e.} if the detector was {\it on} or {\it off}).
\end{enumerate}
\end{framed}

Clearly,  the previous thought experiment constitutes a communication protocol in which Bob can send one bit of information to Alice.
Moreover, for a large enough mass $m$ or for a large enough charge $q$, the test particle of Bob can become entangled with Alice's particle in an arbitrarily short time. But then, apparently, Bob can send a message to Alice faster than light  violating the fundamental principle of relativistic causality. How can we solve this  paradox?
Let us make a list of possible solutions:
\begin{enumerate}
\item[a)] It is impossible to prepare a macroscopic superposition state or to preserve its coherence because of some unknown intrinsic effect lying outside the theory of quantum mechanics.

\item[b)] Once the superposition is created the particle is entangled with the gravitational / electromagnetic field and the reduced state of the particle is mixed. Independently on the measurement of Bob, Alice always finds the state of the particle incoherent unless the entanglement process can be in some way reversed.

\item[c)] Independently on the presence or absence of entanglement, Alice cannot locally discriminate instantaneously whether the superposition is coherent or not. More quantitatively we have that, if Bob is able to generate entanglement in a time $T_{\mathrm{B}}$ and if $T_{\mathrm{A}}$ is the time necessary to Alice for performing her discrimination measurement, then relativistic causality requires
\begin{equation}
T_{\mathrm{A}}+T_{\mathrm{B}} \ge \frac{R}{c}\;. \label{causality}
\end{equation}
Therefore, whenever entanglement can be generated in a time $T_{\mathrm{B}} \le R/c$, we get a non-trivial lower bound on $T_{\mathrm{A}}$.
\end{enumerate}

Anomalous decoherence effects \cite{penrose, GRW, diosi,karolyhazy, review} (as {\it e.g.}\ the Penrose spontaneous localization model) are important open problems in the foundations of quantum mechanics and cannot be excluded  a priori. Up to now however their existence was never experimentally demonstrated and therefore, instead of closing our discussion by directly invoking point a), we try to remain within the framework of quantum mechanics and check if points b) or c) are plausible solutions.

The reader who is familiar with the field of open quantum systems may find the option b) very natural. In standard non-relativistic quantum mechanics, the formation of entanglement between a system and its environment is widely accepted as the origin of any observed form of decoherence. Indeed this approach has also been used to explain the decoherence of moving charged particles, mainly focusing to the double-slit interference experiment~\cite{petruccione}.
It has been recognized by previous works that in a double-slit  experiment there is a limit to the charge of the particle above which photons are emitted due to the acceleration associated to the interference paths~\cite{petruccione,baym}. For large charges then, the particle entangles with the emitted photons and this effect can destroy the  interference pattern.
The reader can then notice that also in our case the particle needs to be accelerated when it is put in the superposition~\eqref{LR}, and if it is charged it will radiate and can become entangled with the emitted photons.
Similarly, an accelerated mass generates gravitational radiation and can become entangled with the emitted gravitons.
However, in Appendix~\ref{rad} we prove that, if the accelerations are slow enough, the resulting quantum state of the electromagnetic field has almost overlap one with the vacuum, and therefore the particle does not get entangled with the emitted photons because no photons at all are emitted.
The same argument can be repeated for the gravitational radiation in the linear approximation.
The reader may now think that the particle in the superposition~\eqref{LR} is entangled at least with its static Coulomb electric field.
However, as we show in details in Appendix~\ref{dof}, the static Coulomb electric field is not a propagating degree of freedom since it vanishes in absence of electric charges.
Thus, the Hilbert space associated to the Coulomb field is the same Hilbert space of the particle, whose reduced state remains pure.

What is then the solution of the paradox?
A reasonable answer appears to be the final option c). Basically, even if the state of the particle is pure and coherent, Alice cannot instantaneously test this fact with a local experiment in her laboratory.
Notice that the hypothesis c) is weaker than the others and can logically coexist with a) and b). Clearly in every case in which a) or b) are valid Alice cannot make any useful experiment because decoherence has already happened.
Therefore we conclude that the weaker and most general solution to the paradox is the fundamental limitation exposed in point c).
In sections~\ref{ent} and~\ref{QED}, we propose two different measurements and show that they are both consistent with this limitation.

\vspace{1 cm}

\section{Minimum discrimination time}
\label{time}
In the previous section we argued that relativistic causality requires a fundamental limitation: Alice's discrimination experiment must be slow. But how slow it has to be? By construction  any thought experiment of the class described before gives a lower bound on the discrimination time $T_{\mathrm{A}}$ whenever $T_{\mathrm{B}} \le R/c$. In what follows we are going to optimize over this class of experiments. We anticipate that this approach leads to the following two bounds which constitute the main results of this work.

\begin{framed}
\noindent{\bf (i) Minimum discrimination time for quantum superpositions of large masses}\\

Given a particle of mass $m$ prepared in a macroscopic quantum superposition of two states separated by a distance $d$, it is impossible to locally discriminate the coherent superposition from an incoherent mixture in a time (up to a multiplicative numerical constant) less than
\begin{equation}
T \simeq \frac{m}{m_{\mathrm{P}}}\,\frac{d}{c}\;,   \label{Tmass}
\end{equation}
where $m_{\rm P} $ is the Planck mass
\begin{equation}
m_{\rm P} = \sqrt{\frac{\hbar c}{G}}  \simeq  2.18 \times 10^{-8} {\rm \; kg}\;.
\end{equation}
\end{framed}

\begin{framed}
\noindent {\bf (ii) Minimum discrimination time for quantum superpositions of large charges}\\

 Given a particle of charge $q$ prepared in a macroscopic quantum superposition of two states separated by a distance $d$, it is impossible to locally discriminate the coherent superposition from an incoherent mixture in a time (up to a multiplicative numerical constant) less than
\begin{equation}
T \simeq \frac{q}{q_{\mathrm{P}}}\,\frac{d}{c}\;,  \label{Tcharge}
\end{equation}
where $q_{\rm P} $ is the Planck charge
\begin{equation}
q_{\rm P} = \sqrt{4 \pi\epsilon_0 \hbar c}  \simeq  11.7\;  e \simeq 1.88 \times 10^{-18} {\rm \; C}\;.
\end{equation}
\end{framed}

Before giving a derivation of the previous results,  we stress that both the bounds \eqref{Tmass} and \eqref{Tcharge} are relevant only for $q \ge q_{\rm P} $ and $m \ge m_{\rm P} $. Indeed for systems below the Planck mass/charge, even if the bounds are formally correct, their meaning is trivial since any measurement of the state must at least interact with both parts of the superposition and this process requires at least a time $d/c$.

\subsection{Dynamics of Bob's test mass}
Let us  first focus on the superpositions of massive particles and give a proof of the bound \eqref{Tmass} (the proof of \eqref{Tcharge} is analogous and will be given later).
It is easy to check that, for a sufficiently narrow trap (detector = {\it off}) the test mass of Bob is insensitive to the gravitational force of Alice's particle and remains stable in its ground state (see Appendix~\ref{strength} for details).
On the contrary, if the trap is removed,  the test mass will experience a different force depending on the position of Alice's particle. The two corresponding Hamiltonians are:
\begin{equation}
\hat{H}_{\rm L}= \frac{\hat{P}^2}{2m_{\mathrm{B}}} - F_{\rm L} \hat{X}\;,   \qquad \hat{H}_{\rm R}=\frac{\hat{P}^2}{2m_{\mathrm{B}}} -F_{\rm R} \hat{X}\;,
\end{equation}
where $m_{\mathrm{B}}$ is the mass of Bob's particle, and $F_{\rm L}$ and $F_{\rm R}$ are the different gravitational forces associated to the ``left" and ``right" positions of  Alice's particle. Their difference
\begin{equation}
\Delta F=  F_{\rm L}-F_{\rm R} \simeq \frac{G\, m_{\mathrm{A}}\, m_{\mathrm{B}}\,  d}{ R^3}\;, \label{deltaFmass}
\end{equation}
where $m_{\mathrm{A}}$ is the mass of Alice's particle, determines the dipole force sensitivity that Bob should be able to detect in order to induce the decoherence of the reduced state possessed by Alice.

Given the initial state of the test mass $|\phi\rangle$, it is easy to check that entanglement can be generated in a time $t$ whenever the different time evolutions associated to $\hat{H}_{\rm L}$ and $\hat{H}_{\rm R}$ drive the test mass into almost orthogonal states, {\it i.e.}
\begin{equation}
\left| \langle \phi | e^{\frac{i}{\hbar} \hat{H}_{\rm R} t }  e^{-\frac{i}{\hbar} \hat{H}_{\rm L} t}    | \phi \rangle \right| \ll 1 . \label{echo}
\end{equation}
Such time depends on the initial state $|\phi \rangle$ and on the Loschmidt echo operator
\begin{equation}\label{Losch}
 \hat{L}(t)=e^{\frac{i}{\hbar} \hat{H}_{\mathrm{R}} t}  e^{-\frac{i}{\hbar} \hat{H}_{\mathrm{L}} t},
\end{equation}
which after two iterations of the Baker-Campbell-Hausdorff formula can be written as
\begin{equation}
 \hat{L}(t)=\exp\left[ i\,\frac{\Delta F}{\hbar} \left(\hat{X}\,t + \frac{\hat{P}}{2m_{\mathrm{B}}}\,t^2   +  \frac{F_1+F_2}{12 m_{\mathrm{B}}}\,  t^3 \right)\right]\;.
\end{equation}
Neglecting the complex phase factor  $e ^{\frac{i}{\hbar}\, \Delta F \, \frac{F_1+F_2}{12 m_{\mathrm{B}}}\,  t^3} $,  $\hat{L}(t)$ is essentially a
phase--space displacement operator of the form  $e ^{\frac{i}{\hbar} (\delta_x  \hat{P} - \delta_p \hat{X})}$,
where
\begin{eqnarray}
\delta_x&=& \frac{\Delta F\, t^2 }{ 2 m_{\mathrm{B}}} ,  \label{deltax} \\
\delta_p&=&-\Delta F \,t\label{deltap}
\end{eqnarray}
are the shifts in position and momentum, respectively.

Since the initial state $|\phi\rangle $ of the test mass is the ground state of a very narrow harmonic trap, it will correspond to a localized Gaussian wave packet which is very noisy in momentum and therefore we may focus only on the position shift  \eqref{deltax} and compare it with the position uncertainty $\Delta X$ of the initial state (see Appendix~\ref{strength} for a detailed proof). We can argue that entanglement is generated only after a time $t=T_{\mathrm{B}}$ such that
\begin{equation}
\frac{\delta x}{\Delta X}= \frac{\Delta F\, T_{\mathrm{B}}^2 }{ 2 m_{\mathrm{B}} \Delta X} \simeq 1. \label{ratio}
\end{equation}

Apparently Bob can generate entanglement arbitrarily quickly by reducing the position uncertainty $\Delta X$. However there is a fundamental limit to the localization precision which is set by the Planck length.
It is widely accepted that no reasonable experiment can overcome this limit~\cite{salecker,mead,hossenfelder}:
\begin{equation}
\Delta X \ge l_{\rm P}= \sqrt{\frac{\hbar G}{c^3}}. \label{minDeltax}
\end{equation}
From Eq.\ \eqref{ratio}, substituting Eq. \eqref{deltaFmass} and using the minimum $\Delta X$ allowed by the constraint \eqref{minDeltax},  we get
\begin{equation}
\frac{\delta x}{\Delta X}=\frac{1}{2}\,\frac{m_{\mathrm{A}}}{m_{\mathrm{P}}}\,\frac{d\   c^{2}   T_{\mathrm{B}}^2}{R^3 }   \simeq 1. \label{ratio2}
\end{equation}
As we have explained in the previous section, relativistic causality implies the inequality \eqref{causality} involving Alice's measurement time $T_{\mathrm{A}}$ and the entanglement time $T_{\mathrm{B}}$. Such inequality provides a lower bound on $T_{\mathrm{A}}$ only if $T_{\mathrm{B}} < R/c$ while it gives no relevant information for $T_{\mathrm{B}} \ge R/c$. Therefore we parametrize $R$ in terms of $T_{\mathrm{B}}$ and a dimensionless parameter $\eta$:
\begin{equation}
 \eta= \frac{c\,T_{\mathrm{B}}}{R}\;, \qquad 0 \le \eta \le 1.
\end{equation}
Using this parametrization, from Eq.\ \eqref{ratio2}, we get

\begin{equation}
T_{\mathrm{B}} \simeq \frac{1}{2}\,\eta^3\,\frac{m_{\mathrm{A}}}{m_{\mathrm{P}}}\,  \frac{d}{ c }  .
\end{equation}
From the causality inequality \eqref{causality} we have
\begin{equation}
T_{\mathrm{A}} + T_{\mathrm{B}} \ge \frac{R}{c} \; \Longrightarrow \; T_{\mathrm{A}} \ge \frac{T_{\mathrm{B}}}{\eta} - T_{\mathrm{B}}= \frac{1}{2}\,\frac{m_{\mathrm{A}}}{m_{\mathrm{P}}}\,\frac{d}{ c }\,(\eta^2- \eta^3).
\end{equation}
Optimizing over $\eta$ we get

\begin{equation}
T_{\mathrm{A}} \ge  \frac{2}{27}\,\frac{m_{\mathrm{A}}}{m_{\mathrm{P}}}\,\frac{d}{c}\;.
\end{equation}
This is, up to a multiplicative numerical constant, the bound given in Eq. \eqref{Tmass}.

\subsection{Dynamics of Bob's test charge}
The calculation in the case in which we have a test charge instead of a test mass is almost identical. The only difference is that Eq. \eqref{deltaFmass} is replaced by the Coulomb counterpart
\begin{equation}
\Delta F=  F_{\rm L}-F_{\rm R} \simeq \frac{q_{\mathrm{A}}\,q_{\mathrm{B}}\,  d}{ 4 \pi \epsilon_0 R^3}\;, \label{deltaFcharge}
\end{equation}
where $q_{\mathrm{A}}$ and $q_{\mathrm{B}}$ are the charges of Alice's and Bob's particles, respectively, while the localization limit \eqref{minDeltax} is replaced by Bob's particle charge radius~\cite{weinberg}
\begin{equation}
\Delta X \ge \frac{q_{\mathrm{B}}}{q_{\mathrm{P}}} \frac{\hbar}{m_{\mathrm{B}} c}. \label{minDeltaxCharge}
\end{equation}
More details on the minimum localization of a macroscopic charge are given in Appendix~\ref{loc}.
From Eq.s \eqref{deltaFcharge} and \eqref{minDeltaxCharge}, repeating exactly the previous argument one finds
\begin{equation}\label{TA}
T_{\mathrm{A}} \ge  \frac{2}{27}\,\frac{q_{\mathrm{A}}}{q_{\mathrm{P}}} \frac{d}{c}\;,
\end{equation}
which is, up to a multiplicative numerical constant, the bound given in Eq. \eqref{Tmass}.

\section{Minimum time from entanglement with radiation}
\label{ent}
In section~\ref{time} we have proved that relativistic causality requires that any measurement Alice can perform to test the coherence of her superposition must require a minimum time, depending on the mass or charge of her particle.
Here we focus on the electromagnetic case, and propose two different measurements to check the coherence of the superposition.

The first is a simplified version of the experiment proposed in Ref.~\cite{scala,wan}.
Let Alice's particle have spin $\frac{1}{2}$, and suppose that her superposition is entangled with the spin, i.e.~\eqref{LR} is replaced by
\begin{equation}
|\psi\rangle=\frac{|L\rangle|\uparrow\rangle+|R\rangle|\downarrow\rangle}{\sqrt{2}}\;.
\end{equation}
Let now Alice apply a spin-dependent force, that vanishes if the spin is up, while brings the particle from $|R\rangle$ to $|L\rangle$ if the spin is down.
If Bob does not perform the measurement, the final state of Alice's particle is
\begin{equation}
\rho_A=|L\rangle\langle L|\otimes|+\rangle\langle +|\;,
\end{equation}
where
\begin{equation}
|+\rangle=\frac{|\uparrow\rangle+|\downarrow\rangle}{\sqrt{2}}\;.
\end{equation}
On the contrary, if Bob induces a collapse of the wave-function, the final state is
\begin{equation}
\rho_A'=|L\rangle\langle L|\otimes\frac{|\uparrow\rangle\langle\uparrow|+|\uparrow\rangle\langle\uparrow|}{2}\;,
\end{equation}
and Alice can test the coherence with a measurement on the spin.

However, this protocol requires the particle to be accelerated if it has spin down and needs to be moved from $|R\rangle$ to $|L\rangle$.
Then it will radiate, and it can entangle with the emitted photons.
A semiclassical computation of the emitted radiation can be found in Appendix~\ref{rad}.
Eq.~\eqref{trad} shows that such radiation is indistinguishable from the vacuum state of the field iff the motion lasts for at least the time required by our previous bound~\eqref{Tcharge}.

\section{Minimum time from quantum vacuum fluctuations}
\label{QED}
In Section~\ref{ent} we have provided an example of experiment able to test the coherence.
The protocol requires to accelerate the charge, and if its duration is too short, the charge radiates and entangles with the emitted photons.
The reader could now think that the bound on the time could be beaten with an experiment that does not involve accelerations.
An example of such experiment could seem to be a measurement of the canonical momentum of Alice's particle.
In this section, we first show that this measurement is indeed able to test the coherence of the superposition
and then we estimate the minimum time necessary to perform it.

The canonical momentum of a charged particle coupled to the electromagnetic field is not gauge invariant, and therefore cannot be directly measured.
Alice can instead measure directly the velocity of her particle, that is gauge invariant.
However, its relation with the canonical momentum now contains the vector potential.
Even if there is no external electromagnetic field, the latter is a quantum-mechanical entity, and is subject to quantum vacuum fluctuations.
Then, the fluctuations of the vector potential enter in the relation between velocity and momentum.
If Alice is not able to measure the field outside her laboratory, she can measure only the velocity of her particle (see Appendix~\ref{dof} for a detailed discussion), and can reconstruct its canonical momentum only if the fluctuations are small.
We show that in an instantaneous measurement these fluctuations are actually infinite.
However, if Alice measures the average of the velocity over a time $T$, they decrease as $1/T^2$, and can be neglected if $T$ is large enough.
This minimum time is found consistent with the bound derived in section~\ref{time}.

In order to simplify our formulas, in this section and in the related Appendices we put as in~\cite{weinberg}
\begin{equation}
\eqbox{\hbar=c=\epsilon_0=\mu_0=1\;,\qquad q_{\mathrm{P}}^2=4\pi\;.}
\end{equation}
These constant will be put back into the final result.

\subsection{The canonical momemtum as a test for coherence}
\label{momentum}
Let us first show that Alice can test the coherence with a measurement of the canonical momentum of her particle.

Let the particle be in the coherent superposition~\eqref{LR} of two identical wave-packets centered in different points, with wave-function
\begin{equation}\label{psiS}
\psi(\mathbf{x})=\frac{\phi(\mathbf{x})+e^{i\varphi}\;\phi(\mathbf{x}-\mathbf{d})}{\sqrt{2}}\;,
\end{equation}
where $\varphi$ is an arbitrary phase.

The probability distribution of the canonical momentum $\hat{P}$ is the modulus square of the Fourier transform of the wave-function:
\begin{equation}\label{osc}
\frac{\left|\psi(\mathbf{k})\right|^2}{(2\pi)^3}=2\cos^2\left(\frac{\mathbf{k}\cdot\mathbf{d}-\varphi}{2}\right)\;\;\frac{\left|\phi(\mathbf{k})\right|^2}{(2\pi)^3}\;,
\end{equation}
and she can test the coherence of the superposition from the interference pattern in momentum space generated by the cosine.
Indeed, an incoherent statistical mixture would be associated to the probability distribution
\begin{equation}
\;\frac{\left|\phi(\mathbf{k})\right|^2}{(2\pi)^3}\;,
\end{equation}
where the cosine squared is replaced by $1/2$, its average over the phase $\varphi$.

Notice from~\eqref{osc} that, in order to be actually able to test the coherence, Alice must measure the canonical momentum with a precision of at least
\begin{equation}\label{prec}
\Delta P\lesssim\frac{\pi}{d}\;,
\end{equation}
where $d=|\mathbf{d}|$.
This precision increases with the separation of the wave-packets, e.g. for $d=1\,\mathrm{m}$, it is $\Delta P\lesssim 10^{-34}\,\mathrm{kg}\cdot \mathrm{m}/\mathrm{s}$.

\subsection{Quantum vacuum fluctuations and minimum time}
Let now Alice's particle carry an electric charge $q$.
We want to take into account the quantum vacuum fluctuations of the electromagnetic field, so quantum electrodynamics is required.
The global Hilbert space is then the tensor product of the Hilbert space of the particle $\mathcal{H}_{\mathrm{A}}$ with the Hilbert space of the field $\mathcal{H}_F$.
The reader can find in Appendix~\ref{emf} the details of the quantization.

The position and canonical momentum operators of Alice's particle $\hat{\mathbf{X}}$ and $\hat{\mathbf{P}}$ still act in the usual way on the particle Hilbert space alone, so that the argument of subsection~\ref{momentum} remains unchanged.
The full interacting Hamiltonian of the particle and the electromagnetic field is
\begin{equation}\label{H}
\hat{H}=\frac{1}{2m}\left(\hat{\mathbf{P}}-q\;\hat{\mathbf{A}}\left(\hat{\mathbf{X}}\right)\right)^2+\hat{H}_F\;,
\end{equation}
where
\begin{equation}\label{AX}
\hat{A}^i\left(\hat{\mathbf{X}}\right)=\int\frac{\hat{a}^i(\mathbf{k})\;e^{i\mathbf{k}\cdot\hat{\mathbf{X}}}+\hat{a}^{i\dag}(\mathbf{k})\;e^{-i\mathbf{k}\cdot\hat{\mathbf{X}}}}{\sqrt{2|\mathbf{k}|}}\;\frac{d^3k}{(2\pi)^3}
\end{equation}
is the vector-potential operator $\hat{\mathbf{A}}(\mathbf{x})$ of~\eqref{Afrx} with the coordinate $\mathbf{x}$ replaced with the position operator $\hat{\mathbf{X}}$, and $\hat{H}_F$ is the free Hamiltonian of the electromagnetic field defined in~\eqref{HF}.

Due to the minimal-coupling substitution, the operator associated to the velocity of the particle is
\begin{equation}\label{Vi}
\hat{\mathbf{V}}\equiv i\left[\hat{H},\;\hat{\mathbf{X}}\right]=\frac{1}{m}\left(\hat{\mathbf{P}}-q\;\hat{\mathbf{A}}\left(\hat{\mathbf{X}}\right)\right)\;,
\end{equation}
that contains the operator vector-potential, and acts also on the Hilbert space of the field.
The canonical momentum can be reconstructed from the velocity with
\begin{equation}\label{PV}
\hat{\mathbf{P}}=m\;\hat{\mathbf{V}}+q\;\hat{\mathbf{A}}\left(\hat{\mathbf{X}}\right)
\end{equation}
if the second term in the RHS can be neglected.
With the help of the commutation relations~\eqref{CCR}, a direct computation of the variance of $\hat{\mathbf{A}}\left(\hat{\mathbf{X}}\right)$ on the vacuum state of the field gives
\begin{equation}\label{A2}
\langle0|{\hat{\mathbf{A}}\left(\hat{\mathbf{X}}\right)}^2|0\rangle=\left(\int\frac{1}{|\mathbf{k}|}\,\frac{d^3k}{(2\pi)^3}\right)\hat{\mathbbm{1}}_{\mathrm{A}}\;,
\end{equation}
that has a quadratic divergence for $\mathbf{k}\to\infty$ due to the quantum vacuum fluctuations.
This divergence can be cured averaging the vector potential over time with a smooth function $\varphi(t)$.
We must then move to the Heisenberg picture, where operators explicitly depend on time, and we define it to coincide with the Schr\"odinger picture at $t=0$, the time at which Alice measures the velocity.
Since the divergence in~\eqref{A2} does not depend neither on the mass nor on the charge of Alice's particle and is proportional to the identity operator on the particle Hilbert space $\hat{\mathbbm{1}}_{\mathrm{A}}$, it has nothing to do with the interaction of the particle with the field.
Then the leading contribution to the result can be computed evolving the field with the free Hamiltonian $\hat{H}_F$ only, i.e. with
\begin{eqnarray}\label{AXt}
&&\hat{A}^i\left(\hat{\mathbf{X}},t\right)=\nonumber\\
&&=\int\frac{\hat{a}^i(\mathbf{k})\;e^{i\left(\mathbf{k}\cdot\hat{\mathbf{X}}-|\mathbf{k}|t\right)}+\hat{a}^{i\dag}(\mathbf{k})\;e^{i\left(|\mathbf{k}|t-\mathbf{k}\cdot\hat{\mathbf{X}}\right)}}{\sqrt{2|\mathbf{k}|}}\;\frac{d^3k}{(2\pi)^3}\;.
\end{eqnarray}
Defining the time-averaged vector potential as
\begin{equation}
\hat{\mathbf{A}}_{av}=\int\hat{\mathbf{A}}\left(\hat{\mathbf{X}},t\right)\;\varphi(t)\;dt\;,
\end{equation}
its variance over the vacuum state of the field is now
\begin{equation}
\langle0|{\hat{\mathbf{A}}_{av}}^2|0\rangle=\left(\frac{1}{2\pi^2}\int_0^\infty \left|\tilde{\varphi}(\omega)\right|^2\omega\;d\omega\right)\hat{\mathbbm{1}}_{\mathrm{A}}\;,
\end{equation}
where
\begin{equation}
\tilde{\varphi}(\omega)=\int\varphi(t)\;e^{i\omega t}\;dt
\end{equation}
is the Fourier transform of $\varphi(t)$.
Taking as $\varphi(t)$ a normalized Gaussian function of width $T$ centered at $t=0$:
\begin{equation}
\varphi(t)=\frac{e^{-\frac{t^2}{2T^2}}}{\sqrt{2\pi}\;T}\;,
\end{equation}
we get as promised a finite result proportional to $1/T^2$:
\begin{equation}\label{DA2}
\langle0|{\hat{\mathbf{A}}_{av}}^2|0\rangle=\frac{\hat{\mathbbm{1}}_{\mathrm{A}}}{4\pi^2 T^2}\;.
\end{equation}
Then, if Alice estimates one component of the canonical momentum (say the one along the $x$ axis) with the time average of the velocity taken with the function $\varphi(t)$, she commits an error of the order
\begin{equation}\label{DeltaPf}
\Delta P\simeq \frac{q}{2\pi\sqrt{3}\;T}\;.
\end{equation}
Comparing~\eqref{DeltaPf} with the required precision to test the coherence~\eqref{prec}, the minimum time required is
\begin{equation}
T\gtrsim\frac{1}{\sqrt{3\pi^3}}\,\frac{q}{q_{\mathrm{P}}}\,\frac{d}{c}\simeq 0.10\,\frac{q}{q_{\mathrm{P}}}\,\frac{d}{c}\;,
\end{equation}
in agreement with the bound~\eqref{TA} imposed by relativistic causality alone.

\section{Conclusions}
In this work we have studied the limitations that the gravitational and electric fields produced by a macroscopic particle impose  on quantum superposition experiments.  We found that, in order to avoid a contradiction between quantum mechanics and relativistic causality, a minimum time is necessary in order to discriminate a coherent superposition from an incoherent statistical mixture. This discrimination time is proportional to the separation distance of the superposition and to the mass (or charge) of the particle.

In the same way as the Heisenberg uncertainty principle inspired the development of a complete theory of quantum mechanics,  our fundamental and quantitative bounds on the discrimination time can be useful for the development of current and future theories of quantum gravity.
Moreover, despite an experimental observation of our results clashes with the difficulty of preparing superpositions of masses above the Planck scale,  the current technological progress on highly massive quantum optomechanical and electromechanical systems provides a promising context \cite{romero-isart, marshall, pikovski, bawaj,schnabel,scala,wan} for testing our predictions.

We thank Seth Lloyd, Leonardo Mazza, and Andrea Tomadin  for fruitful discussions. GdP thanks M. C. Sormani for useful comments.

\newpage
\section*{Supplemental Material}
\appendix
\section{Strength of the trap}
\label{strength}
In this Appendix we show that for Bob a measurement of the position is always better than a measurement of the momentum, i.e. it permits to distinguish the force difference $\Delta F$ in a shorter time.

Let $\omega$ be the frequency of the harmonic trap. The spatial width of its ground state is given by
\begin{equation}\label{psi0}
\Delta X^2\simeq\frac{\hbar}{m_{\mathrm{B}}\,\omega}\;.
\end{equation}
This ground state is insensible to the force difference $\Delta F$ iff the displacement that it generates is less than $\Delta X$, i.e.
\begin{equation}\label{ins}
\frac{\Delta F}{m_{\mathrm{B}}\,\omega^2}\lesssim \Delta X\;.
\end{equation}
Eliminating $\omega$ with~\eqref{psi0}, the inequality~\eqref{ins} becomes hence
\begin{equation}\label{ins2}
\Delta X^3\lesssim\frac{\hbar^2}{m_{\mathrm{B}}\,\Delta F}\;,
\end{equation}
which is the condition we have to enforce to ensure that Bob's detector is ineffective when switched off.
Suppose then that, after switching on the detector,  Bob  tries  to distinguish the two states of Alice by a measurement of $P$: accordingly the momentum spread $\Delta P$ of his initial state must be lower than the displacement in momentum $|\delta_p|=\Delta F\,t$ given by~\eqref{deltap}.
Recalling that  Heisenberg's uncertainty principle $\Delta X\,\Delta P\geq\hbar$ is saturated by a Gaussian pure state, the minimum time after which Bob can distinguish is
\begin{equation}
T_{\mathrm{B}}'=\frac{\hbar}{\Delta F\,\Delta X}\;.
\end{equation}
On the other hand, from~\eqref{ratio} the minimum discrimination time with a measurement of $x$ is
\begin{equation}
T_{\mathrm{B}}=\sqrt{\frac{m_{\mathrm{B}}\,\Delta X}{\Delta F}}\;.
\end{equation}
The reader can check that~\eqref{ins2} implies $T_{\mathrm{B}}\leq T_{\mathrm{B}}'$, i.e. if the trap is strong enough to be insensible to the force difference, for Bob it is always better to measure the position of his particle rather than its momentum.

\section{Maximum localization of a charge}
\label{loc}
In this Appendix we prove that the minimum width over which a charge $q$ greater than the Planck charge $q_{\mathrm{P}}$ can be localized is its charge radius~\eqref{minDeltaxCharge}.

Let us suppose to use a harmonic trap of frequency $\omega$ to localize the charge.
One could think that in principle, with a strong enough trap, the charge can be arbitrarily localized.
However, from the Larmor formula~\cite{jackson} we know that a classical particle with charge $q$ following a harmonic motion of frequency $\omega$ and width $\Delta X$ loses into electromagnetic radiation a power
\begin{equation}\label{larmor}
\frac{dE}{dt}\simeq \frac{q^2\,\omega^4\,\Delta X^2}{\epsilon_0\,c^3}\;.
\end{equation}
In the quantum case, the charge radiates until it gets to the ground state of the trap, where it cannot radiate anymore since there are no other states with a lower energy to go.
However, if the trap is very strong, its ground state is very localized, and therefore has a great uncertainty in velocity.
Since any moving charge generates a magnetic field, this velocity uncertainty generates a large uncertainty in the magnetic field, resulting in a large entanglement between the state of the particle and the state of the field.
Qualitatively, this happens when the energy classically radiated in a period becomes greater than $\hbar\omega$, the energy of the first excited state.
Combining~\eqref{larmor} with~\eqref{psi0}, this happens exactly when the localization $\Delta X$ becomes smaller that the charge radius:
\begin{equation}\label{DX0}
\Delta X\lesssim \frac{q}{q_{\mathrm{P}}}\,\frac{\hbar}{mc}\;.
\end{equation}
Then, if we want the reduced state of the particle to remain pure, we can localize it only up to the limit in~\eqref{DX0}.

\section{Quantization of the electromagnetic field}\label{emf}
We recall here the basics of the quantization of the electromagnetic field.
More details can be found in Ref.~\cite{cohen}.

We denote with $\hat{O}$ an operator in the Schr\"odinger picture, and with $\hat{O}(t)$ its Heisenberg-picture counterpart.
The two pictures are defined to coincide for $t=0$, i.e. $\hat{O}(0)=\hat{O}$.
We recall that in the Heisenberg picture the operators are evolved with the full interacting Hamiltonian.

An Hamiltonian formulation of electrodynamics requires the introduction of the scalar and vector potentials $V$ and $\mathbf{A}$.
It is convenient to Fourier-transform with respect to $\mathbf{x}$.
The potentials are related to the electric and magnetic fields by
\begin{eqnarray}\label{Et}
\hat{\mathbf{E}}(\mathbf{k},t)&=&-i\mathbf{k}\;\hat{V}(\mathbf{k},t)-\frac{\partial}{\partial t}\hat{\mathbf{A}}(\mathbf{k},t)\\
\hat{\mathbf{B}}(\mathbf{k},t)&=&i\mathbf{k}\times\hat{\mathbf{A}}(\mathbf{k},t)\;.\label{Bt}
\end{eqnarray}
We choose the Coulomb gauge, in which the divergence of the vector potential is set to zero at the operator level:
\begin{equation}\label{coulomb}
\mathbf{k}\cdot\hat{\mathbf{A}}(\mathbf{k},t)=0\;.
\end{equation}
It is now convenient to define the ladder operators
\begin{equation}\label{ai}
\hat{a}^i(\mathbf{k},t)\equiv\sqrt{\frac{|\mathbf{k}|}{2}}\;\hat{A}^i(\mathbf{k},t)+\frac{i}{\sqrt{2|\mathbf{k}|}}\;\frac{\partial}{\partial t}\hat{A}^i(\mathbf{k},t)\;,
\end{equation}
satisfying the constraint $k_i\,\hat{a}^i(\mathbf{k},t)=0$ as a consequence of~\eqref{coulomb}.
The definition in~\eqref{ai} can be inverted:
\begin{equation}\label{Afrx}
\hat{A}^i(\mathbf{x},t)=\int\frac{\hat{a}^i(\mathbf{k},t)\;e^{i\mathbf{k}\cdot\mathbf{x}}+\hat{a}^{i\dag}(\mathbf{k},t)\;e^{-i\mathbf{k}\cdot\mathbf{x}}}{\sqrt{2|\mathbf{k}|}}\;\frac{d^3k}{(2\pi)^3}\;.
\end{equation}
The ladder operators satisfy the equal-time canonical commutation relations
\begin{eqnarray}\label{CCR}
\left[\hat{a}^i(\mathbf{k},t),\;\hat{a}^{j\dag}(\mathbf{q},t)\right]&=&\Pi^{ij}(\mathbf{k})\;(2\pi)^3\delta^3(\mathbf{k}-\mathbf{q})\\
\left[\hat{a}^i(\mathbf{k},t),\;\hat{a}^{j}(\mathbf{q},t)\right]&=&\left[\hat{a}^{i\dag}(\mathbf{k},t),\;\hat{a}^{j\dag}(\mathbf{q},t)\right]=0\;, \quad
\end{eqnarray}
where $\Pi(\mathbf{k})$ is the projector onto the subspace orthogonal to $\mathbf{k}$:
\begin{equation}
\Pi^{ij}(\mathbf{k})=\delta^{ij}-\frac{k^i\,k^j}{\mathbf{k}^2}\;.
\end{equation}
The vacuum state of the field $|0\rangle$ is defined as the state annihilated by all the Schr\"odinger-picture annihilation operators:
\begin{equation}\label{vac}
\hat{a}^i(\mathbf{k})|0\rangle=0\qquad\forall\;\mathbf{k}\in\mathbb{R}^3\;,\quad i=1,\,2,\,3\;,\qquad|0\rangle\in\mathcal{H}_F\;.
\end{equation}
Besides, the $\hat{a}^i(\mathbf{k})$ together with their hermitian conjugates $\hat{a}^{i\dag}(\mathbf{k})$ generate the whole observable algebra of $\mathcal{H}_F$.

Maxwell's equations determine the time evolution of the ladder operators:
\begin{equation}
\frac{\partial}{\partial t}\hat{a}^i(\mathbf{k},t)+i|\mathbf{k}|\,\hat{a}^i(\mathbf{k},t)=\frac{i\,\Pi^i_{\phantom{i}j}(\mathbf{k})}{\sqrt{2|\mathbf{k}|}}\hat{J}^i(\mathbf{k},t)\;,\label{Ai}
\end{equation}
where $\hat{\mathbf{J}}$ is the operator associated to the current density of the quantum system interacting with the electromagnetic field.
Eq.~\eqref{Ai} is easily integrated:
\begin{align}\label{ait}
&\hat{a}^i(\mathbf{k},t)=\nonumber\\
&=e^{-i|\mathbf{k}|t}\left(\hat{a}^i(\mathbf{k})+\frac{i\,\Pi^i_{\phantom{i}j}(\mathbf{k})}{\sqrt{2|\mathbf{k}|}}\;\int_0^t e^{i|\mathbf{k}|t'} \hat{J}^i(\mathbf{k},t')\;dt'\right)\;,
\end{align}
where we have imposed the Heisenberg and Schr\"odinger pictures to coincide at $t=0$, i.e. $\hat{a}^i(\mathbf{k},0)=\hat{a}^i(\mathbf{k})$.

In the free case, i.e. when the current vanishes at the operator level ($\hat{\mathbf{J}}(\mathbf{k},t)=0$), the relation between the two pictures is given by the free Hamiltonian
\begin{equation}\label{HF}
\hat{H}_F\equiv\int|\mathbf{k}|\;\hat{a}_i^\dag(\mathbf{k})\;\hat{a}^i(\mathbf{k})\;\frac{d^3k}{(2\pi)^3}\;,
\end{equation}
i.e.
\begin{equation}\label{afree}
\hat{a}^i(\mathbf{k},t)=e^{-i|\mathbf{k}|t}\hat{a}^i(\mathbf{k})=e^{i\hat{H}_Ft}\;\hat{a}^i(\mathbf{k})\;e^{-i\hat{H}_Ft}\;.
\end{equation}

\section{Coherent states}
\label{coh}
We introduce now the set of coherent states of the electromagnetic field.
Our formalism is analogue to the one of Refs.~\cite{BR}, where the reader is referred for further details.

For any function $f:\mathbb{R}^3\to\mathbb{C}^3$ subject to the constraint
\begin{equation}
k_i\,f^i(\mathbf{k})=0\qquad\forall\;\mathbf{k}\in\mathbb{R}^3\;,
\end{equation}
define the unitary displacement operator
\begin{align}
&\hat{D}[f]\equiv\exp\left(\int\left(f_i(\mathbf{k})\,\hat{a}^{i\dag}(\mathbf{k})-f_i^*(\mathbf{k})\,\hat{a}^i(\mathbf{k})\right)\frac{d^3k}{(2\pi)^3}\right)\;,\\
&\hat{D}^\dag[f]=\hat{D}[-f]\;,
\end{align}
acting on the ladder operators as
\begin{equation}\label{dispa}
\hat{D}^\dag[f]\;\hat{a}^i(\mathbf{k})\;\hat{D}[f]=\hat{a}^i(\mathbf{k})+f^i(\mathbf{k})\;.
\end{equation}
Their composition rule is
\begin{eqnarray}\label{UU}
&&\hat{D}[f]\;\hat{D}[g]=\hat{D}[f+g] \times \;  \nonumber \\
&&\times\exp{\left(\frac{1}{2}\int\left(f^i(\mathbf{k})\,g_i^*(\mathbf{k})-f_i^*(\mathbf{k})\,g^i(\mathbf{k})\right)\frac{d^3k}{(2\pi)^3}\right)}\;. \qquad
\end{eqnarray}
We can now define the coherent states with a displacement operator acting on the vacuum state of the field:
\begin{equation}
|f\rangle\equiv\hat{D}[f]|0\rangle\in\mathcal{H}_F\;,
\end{equation}
that are eigenstates of the annihilation operators:
\begin{equation}
\hat{a}^i(\mathbf{k})|f\rangle=f^i(\mathbf{k})|f\rangle\;.
\end{equation}
Their overlap is
\begin{equation}\label{overlap}
\left|\langle f|g\rangle\right|^2=\exp\left(-\int\left|f(\mathbf{k})-g(\mathbf{k})\right|^2\frac{d^3k}{(2\pi)^3}\right)\;.
\end{equation}

\section{Radiation emitted by Alice's particle}
\label{rad}
In this Appendix we consider a classical charged particle following an accelerated trajectory and coupled to the quantum electromagnetic field.
We show that the radiation classically emitted by the particle induces on the quantum field a displacement operator, and if the initial state is the vacuum, it is brought to a coherent state (see Appendix~\ref{coh} for the definition).
It turns out that, if the accelerations of the particle are smooth enough, this coherent state has overlap almost one with the vacuum, i.e. it is not distinguishable from it.
Then, for a slow motion the particle does not radiate photons at all, and therefore it does not get entangled with the field.

Let $\mathbf{J}$ be the classical current density associated to the trajectory of the particle.
Looking at the time evolution equation for the ladder operators~\eqref{ait}, and recalling~\eqref{dispa} and~\eqref{afree}, it is easy to show that such evolution is provided by a displacement operator, i.e.
\begin{equation}
\hat{a}^i(\mathbf{k},t)=\hat{D}^\dag[f]\;e^{i\hat{H}_Ft}\;\hat{a}^i(\mathbf{k})\;e^{-i\hat{H}_Ft}\;\hat{D}[f]\;,
\end{equation}
where
\begin{equation}\label{falpha}
f^i(\mathbf{k})=\frac{i\,\Pi^i_{\phantom{i}j}(\mathbf{k})}{\sqrt{2|\mathbf{k}|}}\;\int_0^t e^{i|\mathbf{k}|t'} J^i(\mathbf{k},t')\;dt'\;.
\end{equation}
Since the ladder operators generate the whole observable algebra of $\mathcal{H}_F$, if the fields starts in the vacuum, its time-evolved state is the coherent state $e^{-i\hat{H}_Ft}|f\rangle$.
Its overlap with the vacuum can be computed with~\eqref{overlap}:
\begin{equation}\label{cab}
\left|\langle 0|f\rangle\right|^2=\exp\left(-\int\left|f(\mathbf{k})\right|^2\frac{d^3k}{(2\pi)^3}\right)\;.
\end{equation}

We now consider a point particle carrying charge $q$ that starts in $\mathbf{x}=\mathbf{0}$ at $t=0$, and in a time $t_0$ is brought to the position $\mathbf{x}=\mathbf{d}$ with a trajectory described by $\mathbf{x}(t)$.
The current density is then
\begin{equation}\label{Jp}
\mathbf{J}(\mathbf{k},t)=q\;\mathbf{v}(t)\;e^{-i\mathbf{k}\cdot\mathbf{x}(t)}\;,
\end{equation}
where $\mathbf{v}(t)\equiv\frac{d}{dt}\mathbf{x}(t)$ is the particle velocity.
For wavelengths large with respect to the extension of the motion, i.e. for
\begin{equation}\label{longw}
|\mathbf{k}|\ll\frac{1}{d}\;,
\end{equation}
the phase factor in~\eqref{Jp} can be discarded, getting
\begin{equation}
\mathbf{J}(\mathbf{k},t)\simeq q\;\mathbf{v}(t)\;.
\end{equation}
We want to look at the state of the field after the particle has reached the new position $\mathbf{x}=\mathbf{d}$, i.e. for $t>t_0$.
Since the velocity $\mathbf{v}(t)$ vanishes for $t\leq0$ and $t\geq t_0$, the displacement of~\eqref{falpha} becomes
\begin{equation}\label{fq}
f^i(\mathbf{k})=\frac{i\,q}{\sqrt{2|\mathbf{k}|}}\;\Pi^i_{\phantom{i}j}(\mathbf{k})\;v^j(\omega=|\mathbf{k}|)\;,
\end{equation}
where
\begin{equation}
\mathbf{v}(\omega)=\int_{-\infty}^\infty \mathbf{v}(t)\;e^{i\omega t}\;dt
\end{equation}
is the Fourier transform of the velocity.
Putting~\eqref{fq} into~\eqref{cab}, the overlap becomes
\begin{equation}\label{ovw}
\left|\langle0|f\rangle\right|^2=\exp\left(-\frac{q^2}{6\pi^2}\int_0^\infty|\mathbf{v}(\omega)|^2\;\omega\;d\omega\right)\;.
\end{equation}
For simplicity we consider a one-dimensional motion, and we put the $x$ axis in the direction of $\mathbf{d}$.
As an example, we take
\begin{equation}
x(t)=d\;\sin^2\left(\frac{\pi}{2}\,\frac{t}{t_0}\right)\qquad\text{for}\;0\leq t\leq t_0\;,
\end{equation}
satisfying the conditions
\begin{equation}
x(0)=0\;,\qquad x(t_0)=d\;,\qquad v(0)=v(t_0)=0\;.
\end{equation}
The Fourier transform of the velocity is
\begin{equation}\label{wif}
v(\omega)=e^\frac{i\omega t_0}{2}\;\frac{d\cos\frac{\omega t_0}{2}}{1-\frac{\omega^2t_0^2}{\pi^2}}\;,
\end{equation}
and the overlap
\begin{eqnarray}\label{result}
\left|\langle0|f\rangle\right|^2 &=&\exp\left(-\pi\,\frac{\pi\mathrm{Si}(\pi)-2}{6}\;\frac{q^2}{q_{\mathrm{P}}^2}\;\frac{d^2}{c^2\,t_0^2}\right) \nonumber\\
&\simeq&\exp\left(-2\;\frac{q^2}{q_{\mathrm{P}}^2}\;\frac{d^2}{c^2\,t_0^2}\right)\;,
\end{eqnarray}
where $\mathrm{Si}(x)$ is the sine integral function
\begin{equation}
\mathrm{Si}(x)\equiv\int_0^x\frac{\sin y}{y}dy\;.
\end{equation}
Looking at~\eqref{wif}, the dominant contribution to the integral in~\eqref{ovw} comes from the region $\omega\,t_0\lessapprox1$.
The approximation in~\eqref{longw} is then valid iff $d\ll c\,t_0$, i.e. if the motion is not relativistic.

The final result~\eqref{result} tells us that for a fixed distance $d$, no photons are radiated if the motion lasts for at least
\begin{equation}\label{trad}
t_0\gtrsim \sqrt{2}\;\frac{q}{q_{\mathrm{P}}}\;\frac{d}{c}\;.
\end{equation}
Then, Alice can always create the coherent superposition~\eqref{LR} without entangling her particle with the emitted photons provided she has enough time to do it.
Besides, if Alice wants to perform the measurement of Section~\ref{ent}, she needs at least a time~\eqref{trad} to bring the state $|R\rangle$ back to $|L\rangle$ if she does not want to entangle with the emitted photons.

\section{Absence of entanglement with the static electric field}
\label{dof}
In this Appendix we explain in detail why when Alice's charged particle is in the superposition~\eqref{LR}, despite it generates a static electric field that depends on its position, it is still not entangled with the field, and the global state is a product with the field part in the vacuum.

The first Maxwell's equation reads
\begin{equation}\label{A0}
\mathbf{k}^2\hat{V}(\mathbf{k})=\hat{\rho}(\mathbf{k})\;,
\end{equation}
and completely determines the electric potential operator $\hat{V}$ in terms of the charge density operator $\hat{\rho}$:
\begin{equation}\label{A0J2}
\hat{V}(\mathbf{k})=\frac{1}{\mathbf{k}^2}\,\hat{\rho}(\mathbf{k})\;.
\end{equation}
Putting together~\eqref{A0J2} and~\eqref{Et}, the electric field is given by
\begin{equation}\label{EA}
\hat{\mathbf{E}}(\mathbf{k})=-\frac{i\mathbf{k}}{\mathbf{k}^2}\;\hat{\rho}(\mathbf{k})-\frac{\partial}{\partial t}\hat{\mathbf{A}}(\mathbf{k})\;.
\end{equation}
Then, the longitudinal (i.e. proportional to $\mathbf{k}$) component of the electric field operator is determined by the charge-density operator, and acts on the Hilbert space of the particle alone.
Therefore, even if the field is in its vacuum state~\eqref{vac}, the expectation value of the electric field is the static Coulomb electric field generated by the expectation value of the charge density, and hence depends on the particle wave-function.
This means that the state of the field alone does not contain all the information on the electric field, since its longitudinal component is encoded into the state of the particle.

Seen from a different perspective, the longitudinal component of the electric field is not a dynamical propagating degree of freedom, since it vanishes in absence of external charges and is completely determined by them, so there is no Hilbert space associated to it.
The Hilbert space of the field contains only the degrees of freedom associated to the electromagnetic radiation, i.e. the magnetic field and the transverse (orthogonal to $\mathbf{k}$) component of the electric field.
Then in a product state with the field part in the vacuum only these components are in the vacuum mode, while there can be a static electric field depending on the state of the particle.

\section{Locality}
In this Appendix we explain in detail why Alice can measure only the velocity of her particle, and not its canonical momentum, if she is constrained to remain in her laboratory, which has the size of the support of the wave-function of the particle.

The wave-function $\psi(\mathbf{x},t)$ of a particle carrying electric charge $q$ coupled to an electromagnetic field is invariant under the joint gauge transformation~\cite{cohen}
\begin{eqnarray}
\psi'(\mathbf{x},t) &=& e^{iq\Lambda(\mathbf{x},t)}\;\psi(\mathbf{x},t)\\
\mathbf{A}'(\mathbf{x},t) &=& \mathbf{A}(\mathbf{x},t)+\nabla\Lambda(\mathbf{x},t)\label{deltaA}\\
{V}'(\mathbf{x},t) &=& V(\mathbf{x},t)-\frac{\partial}{\partial t}\Lambda(\mathbf{x},t)\;.
\end{eqnarray}
The canonical momentum $\hat{\mathbf{P}}=-i\nabla$ is not gauge invariant, but transforms in the Heisenberg picture as
\begin{equation}\label{deltaP}
\hat{\mathbf{P}}'(t)=\hat{\mathbf{P}}(t)+q\nabla\Lambda\left(\hat{\mathbf{X}}(t),t\right)\;,
\end{equation}
and therefore Alice cannot measure it directly.
The reader can easily check using~\eqref{deltaA} and~\eqref{deltaP} that the velocity given by~\eqref{Vi} is gauge invariant, as it has to be.
Alice can then measure directly the velocity, and reconstruct from it the canonical momentum.
However, the relation between them (Eq.~\eqref{PV}) contains the vector potential, that from~\eqref{deltaA} is not gauge invariant, and cannot be directly measured.
In the Coulomb gauge, it is possible to invert~\eqref{Bt} and express the vector potential in terms of the magnetic field, that is gauge invariant and can be actually measured by Alice:
\begin{equation}\label{AB}
\hat{\mathbf{A}}(\mathbf{x})=\frac{1}{4\pi}\int\frac{\nabla\times\hat{\mathbf{B}}(\mathbf{y})}{\left|\mathbf{x}-\mathbf{y}\right|}\;d^3y\;.
\end{equation}
Putting together~\eqref{AB} and~\eqref{PV}, we get
\begin{equation}\label{PVB}
\hat{\mathbf{P}}=m\;\hat{\mathbf{V}}+\frac{q}{4\pi}\int\frac{\nabla\times\hat{\mathbf{B}}(\mathbf{y})}{\left|\hat{\mathbf{X}}-\mathbf{y}\right|}\;d^3y\;.
\end{equation}
However, reconstructing the canonical momentum from the velocity with~\eqref{PVB} requires Alice to measure the magnetic field in the whole space.
Even if she can allow for some error in the reconstruction, the region in which she has to measure the field increases with the charge $q$, and can extend well outside the support of the wave-function.

\end{document}